Multi-View Variational Autoencoder for Missing Value Imputation in Untargeted Metabolomics

Chen Zhao[1#], Kuan-Jui Su[2#], Chong Wu[3], Xuewei Cao[4], Qiuying Sha[4], Wu Li[2], Zhe Luo[2], Tian Qin[2], Chuan Qiu[2], Lan Juan Zhao[2], Anqi Liu[2], Lindong Jiang[2], Xiao Zhang[2], Hui Shen[2], Weihua Zhou[1,5*], Hong-Wen Deng[2*]

1. Department of Applied Computing, Michigan Technological University, 1400 Townsend Dr, Houghton, MI, 4993

2. Division of Biomedical Informatics and Genomics, Tulane Center of Biomedical Informatics and Genomics, Deming Department of Medicine, Tulane University, New Orleans, LA 70112

3. Department of Biostatistics, University of Texas MD Anderson, Pickens Academic Tower, 1400 Pressler St., Houston, TX 77030

4. Department of Mathematical Sciences, Michigan Technological University, 1400 Townsend Dr, Houghton, MI, 49931

5. Center for Biocomputing and Digital Health, Institute of Computing and Cybersystems, and Health Research Institute, Michigan Technological University, Houghton, MI 49931

# Chen Zhao and Kuan-Jui Su contribute equally.

* Corresponding authors and lead contacts:

Hong-Wen Deng, Ph.D.

Tulane Center for Biomedical Informatics and Genomics, Deming Department of Medicine, Tulane University,

1440 Canal Street, Suite 1619F, New Orleans, LA 70112, USA

Tel: 504-988-1310

Email: hdeng2@tulane.edu

Weihua Zhou, Ph.D.

Department of Applied Computing, Michigan Technological University,

1400 Townsend Dr, Houghton, MI, 49931, USA

Tel: 906-487-2666

E-Mail: whzhou@mtu.edu

**Abstract**

*Background*: Missing data is a common challenge in mass spectrometry-based metabolomics, which can lead to biased and incomplete analyses. The integration of whole-genome sequencing (WGS) data with metabolomics data has emerged as a promising approach to enhance the accuracy of data imputation in metabolomics studies.

*Method*: In this study, we propose a novel method that leverages the information from WGS data and reference metabolites to impute unknown metabolites. Our approach utilizes a multi-view variational autoencoder to jointly model the burden score, polygenetic risk score (PGS), and linkage disequilibrium (LD) pruned single nucleotide polymorphisms (SNPs) for feature extraction and missing metabolomics data imputation. By learning the latent representations of both omics data, our method can effectively impute missing metabolomics values based on genomic information.

*Results*: We evaluate the performance of our method on empirical metabolomics datasets with missing values and demonstrate its superiority compared to conventional imputation techniques. Using 35 template metabolites derived burden scores, PGS and LD-pruned SNPs, the proposed methods achieved $R^2$-scores > 0.01 for 71.55% of metabolites.

*Conclusion*: The integration of WGS data in metabolomics imputation not only improves data completeness but also enhances downstream analyses, paving the way for more comprehensive and accurate investigations of metabolic pathways and disease associations. Our findings offer valuable insights into the potential benefits of utilizing WGS data for metabolomics data imputation and underscore the importance of leveraging multi-modal data integration in precision medicine research.

# 1. Introduction

Metabolomics is a scientific field that involves the systematic identification and quantification of a broad spectrum of small molecule metabolites present in biological samples, such as cells, tissue, and biological fluids [1]. Mass spectrometry (MS) is a significant high-throughput analytical technique utilized for profiling small molecular compounds, including metabolites, in biological samples [2,3]. The missing values in MS-based metabolomic data are often presented and challenging to handle [4,5], leading to a bias for the downstream analysis [6]. For downstream analysis using metabolomics data, a complete dataset is preferred and often required.

Many machine learning methods have been applied to impute within-omics metabolomics, such as k-nearest neighbors (KNN) imputation [7] and random forest regression (RF) [8]. However, existing within-omics imputation suffers from low accuracy in empirical practice. In addressing this limitation, the practicality of cross-omics based imputation becomes evident. The development of high-throughput omics technologies has revolutionized our ability to study biological systems at a molecular level [9]. These high-throughput techniques, including genomics, transcriptomics, proteomics, and epigenomics, allow us to profile the genetic expression and interaction of molecules from different biological perspectives [10]. In a recent comprehensive analysis using whole-genome sequencing (WGS), it was shown that blood metabolites display a high degree of heritability and consistency [11]. Discovering how genetic variants impact metabolites can provide valuable insights into the molecular mechanisms that influence the development of diseases. This positioning of metabolites along the pathway between genetic determinants and various health outcomes is significant [12]. Integrating these two disparate datasets has the potential to unlock invaluable information, facilitating a deeper understanding of missing value recovery and imputation. Using WGS data as a reference to perform cross-omics imputation for metabolomics data has garnered significant attention [13] for its ability to leverage genetic information in predicting metabolite abundances.

In this study, we propose a novel multi-view variational autoencoder (MVAE) framework for imputing missing values in metabolomics data, leveraging genetic information from WGS data. The workflow of the proposed approach is shown in Figure 1. Our method integrates multiple features, including burden scores from template metabolites, polygenic risk scores (PGS), and linkage disequilibrium (LD)-pruned single nucleotide polymorphisms (SNPs), for comprehensive feature extraction. By fusing information from both WGS and template metabolomics data, our approach achieves cross-omics imputation, enabling a more holistic understanding of the metabolic landscape.

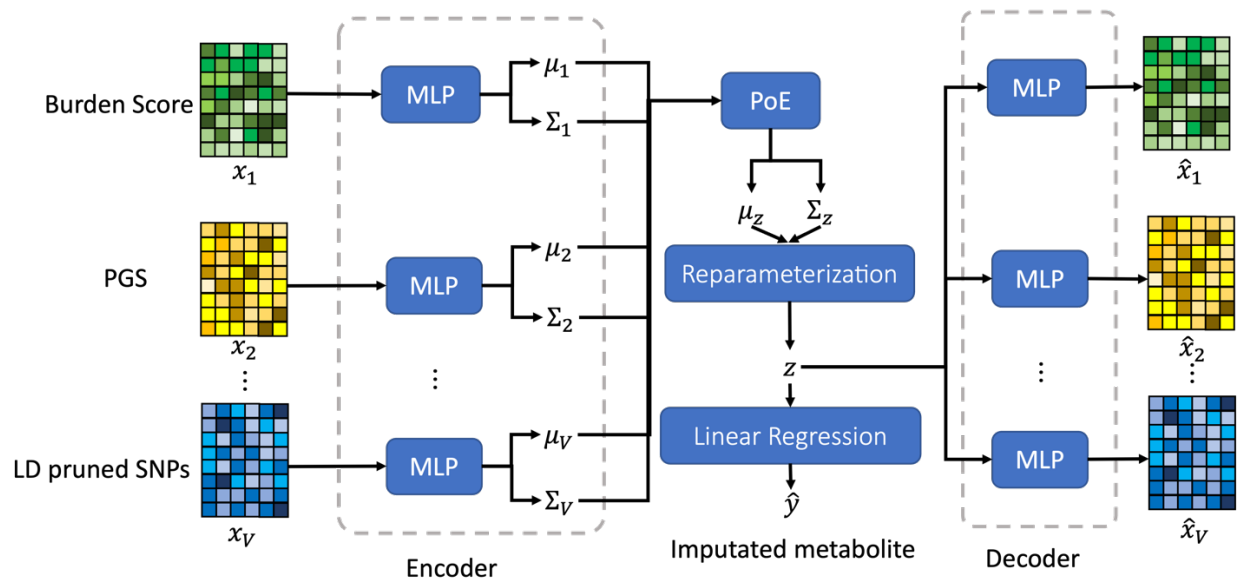

**Figure 1**. The architecture of the proposed MVAE for metabolomics data imputation using burden score, PGS, and LD pruned SNPs. MLP: multi-layer perceptron; PoE: product of experts.

## 2. Materials and methods

### 2.1. Enrolled subjects

The studied cohort was acquired from the Louisiana Osteoporosis Study (LOS) [14,15]. The LOS cohort is an ongoing research dataset (>17,000 subjects accumulated so far with recruitment starting in 2011), aimed at investigating both environmental and genetic risk factors for osteoporosis and other musculoskeletal diseases [16,17]. All participants signed an informed-consent document before any data collection, and the study was approved by the Tulane University Institutional Review Board. A total of 1,110 subjects with both WGS and metabolomics data were enrolled. The demographical information is shown in Table 1.

Table 1. Demographic and Physical Characteristics of Participants (N=1,110)

| Metric | Overall | Stratified by Sex | | Stratified by Race | |
|---|---|---|---|---|---|
| | | Female | Male | African American | White |
| Number of participants (n) | 1110 | 126 | 984 | 418 | 692 |
| Sex = Male (%) | 984 (88.6%) | | | 387 (92.6) | 597 (86.3) |
| Race = White (%) | 692 (62.3%) | 95 (75.4) | 597 (60.7) | | |
| Exercise = TRUE (%) | 823 (74.1%) | 87 (69.0) | 736 (74.8) | 279 (66.7) | 544 (78.6) |
| Age (years)(mean (SD)) | 38.80 (10.87) | 51.99 (14.07) | 37.11 (9.10) | 40.13 (9.26) | 38.00 (11.66) |
| Height (cm) (mean (SD)) | 173.87 (7.78) | 163.45 (5.96) | 175.21 (6.93) | 173.99 (7.53) | 173.80 (7.93) |
| Weight (kg) (mean (SD)) | 81.72 (17.03) | 72.57 (17.45) | 82.89 (16.62) | 82.69 (17.45) | 81.13 (16.76) |

Note: SD = Standard Deviation.

The detailed procedure for WGS has been described elsewhere [18]. Briefly, the WGS of the human peripheral blood DNA was performed with an average read depth of 22′ using a BGISEQ-500 sequencer

(BGI Americas Corporation, Cambridge, MA, USA) of 350 bp paired-end reads [17]. The aligned and cleaned WGS data were mapped to the human reference genome (GRCh38/hg38) using Burrows-Wheeler Aligner software [19] following the recommended best practices for variant analysis with the Genome Analysis Toolkit (GATK) to ensure accurate variant calling. Genomic variations were detected by the HaplotypeCaller of GATK, and the variant quality score recalibration method was applied to obtain high-confidence variant calls [20].

This study employed the liquid chromatography-mass spectrometry (LC-MS) metabolomics platform developed by Metabolon, Inc. (Durham, NC, USA), where they were stored at −80 °C until analysis. All samples were prepared according to the manufacturer's protocol using the automated MicroLab STAR® system (Hamilton, USA). Proteins was precipitated using methanol under vigorous shaking for 2 min (Glen Mills GenoGrinder 2000), followed by centrifugation to recover chemically diverse metabolites. The extracts were then used as input to Waters ACQUITY ultra-performance liquid chromatography (UPLC) and a Thermo Scientific Q-Exactive high resolution/accurate MS interface with a heated electrospray ionization (HESI-II) source and Orbitrap mass analyzer operated at 35,000 mass resolution for positive and negative electrospray ionization. The process details have been described in prior studies [21,22]. We implemented rigorous quality control measures, including the use of a pooled matrix sample as a technical replicate, extracted water samples as process blanks, and the addition of a carefully selected QC standards cocktail to each sample. These measures ensured instrument performance monitoring, aided chromatographic alignment, and minimized interference. Instrument and process variability were assessed through median relative standard deviation calculations. Furthermore, we randomized experimental samples, eliminating biases and ensuring data reliability for all endogenous metabolites present in 100% of the pooled matrix samples.

## 2.2. Data processing

*WGS data processing*. There were a total of 10,623,292 SNPs in the cohort with 1110 subjects. For quality control, we removed genetic variants with missing rates larger than 5% and Hardy-Weinberg equilibrium exact test p-values less than $10^{-4}$. Due to evolutionary dynamics, certain SNPs frequently exhibit variations in a population (referred to as "common" variants), while other SNPs remain identical in the vast majority of the population, with only a few individuals showing mutations (referred to as "rare" variants) - resulting in a form of class imbalance. In this study, we used the minor allele frequency (MAF) of 5% as the cut-off threshold to determine the common and rare variants in our following analyses. Polygenic risk scores, burden scores, and raw SNPs represent distinct genetic modalities that collectively provide a comprehensive view of an individual's genetic predisposition, each contributing unique insights. Thus, we explored three different methods to encode the genetic modalities, including PGS, burden scores, and LD-pruned SNPs.

1. The polygenic score (PGS) is a quantitative measure to estimate an individual's genetic risk to a specific trait or disease [23]. It is calculated as the weighted summation of the genetic variants, where the weights are based on their effect sizes to the trait of interest. The PGS represents the combined genetic risk across common or less common variants [24]. Since PGS offers several clinical benefits, including disease risk prediction, diagnosis, and prognosis [25], we employed the "pgsc_calc" workflow from the PGS Catalog to compute PGS scores using our in-house WGS data. Additionally, we associated our genetic variations with 3,335 predictive traits and diseases, introducing them as new input data [24]. The top 512 PGS with the highest variance were selected as the PGS features for each subject.
2. LD pruned SNPs. LD pruning is a method used to remove redundant genetic variants from a dataset to reduce the effects of LD [26]. The pruning method scans the pairwise correlated SNPs and kept the one with higher MAF. After performing LD pruning for the common variants, 266,240 SNPs were retained and the top 1,024 SNPs with the highest variance were used as the SNPs features. The LD pruning was performed using a window size of 50 with a step size of 5 and a pairwise $R^2$ threshold of 0.5.

3. Burden score. We considered rare variants separately using a widely used burden score [27]. Briefly, for each metabolite, we first regressed metabolite abundance level on the first two genetic principal components to adjust potential population stratification. Second, the burden score was calculated as the summation of each metabolite residual multiplied by the allele count of each rare variant across the genome individually. The top 512 burden score features with the highest variance were selected as a new genetic modality. Burden scores are particularly effective in aggregating rare genetic variants within a specific genomic region or gene. Instead of analyzing each rare variant individually, which might require a large sample size to detect associations, burden scores group these variants together based on their collective impact, which effectively improves the performance of the metabolomics data imputation.

Note that the burden score was calculated with the involvement of metabolites. To avoid using output values as input in our method, we created a new template metabolite that utilizes highly correlated metabolites as a new dependent variable and prior knowledge for model training and testing. In detail, we employ a template set containing $M$ metabolites as the template metabolites. During the model training, the Pearson correlation between a given predicted metabolite and each metabolite in the template set among the enrolled subjects was calculated. The metabolite in the template set with the highest correlation was selected as the template metabolite to calculate the burden score with the selected rare variants.

As a result, three views were obtained to characterize the genetic information, including the PGS scores, LD pruned SNPs, and burden scores.

*Metabolomics data processing.* In the metabolomics profile, we identified 1,839 metabolites. To prove the concept of our developed model, we selected 497 metabolites with missing rate < 5% in our study. To validate and benchmark our proposed method, the missing values were excluded during the experiments and the subjects with the presented metabolites were included.

### 2.3. MVAE

To enhance the clarity of notation, especially regarding vectors, scalars, and their impact on mathematically derived relations, we use ***italic bolded*** font for vector, **bolded** for matrix, and *italic* for scalar.

Before introducing multi-view variational autoencoder, we first introduce the variational autoencoder (VAE). VAE was proposed by Kingma et al. [28], is a latent variable generative model which learns the deep representation of the input data. The goal of VAE is to maximize the marginal likelihood of the data (a.k.a evidence), which can be decomposed into a sum over marginal log-likelihoods of individual features, as illustrated in Eq. 1.

$$\log p_\theta(\boldsymbol{X}^{(i)}) = D_{KL}\left(q_\phi(\boldsymbol{Z}|\boldsymbol{X}^{(i)}) \parallel p_\theta(\boldsymbol{Z}|\boldsymbol{X}^{(i)})\right) + \mathcal{L}(\theta, \phi; \boldsymbol{X}^{(i)}) \qquad (1)$$

where $\boldsymbol{X}^{(i)}$ is the feature vector for $i$-th subject in the dataset $\{\boldsymbol{X}^{(i)}\}_{i=1}^{N}$, $N$ is the number of subjects, $\boldsymbol{Z}$ is a random variable in the latent space, $q_\phi$ is the posterior approximation of $\boldsymbol{Z}$ with the learnable parameters $\phi$, $p_\theta$ is the ground truth posterior distribution of $\boldsymbol{Z}$ with the intractable parameters $\theta$, and $D_{KL}(\cdot\|\cdot)$ represents the Kullback–Leibler (KL) divergence between the approximated posterior distribution and the ground truth posterior distribution. Because of the non-negativity of the KL divergence, the log-likelihood $\log p_\theta(\boldsymbol{X}^{(i)}) \geq \mathcal{L}(\theta, \phi; \boldsymbol{X}^{(i)})$. If the approximated posterior distribution $q_\phi(\boldsymbol{Z}|\boldsymbol{X}^{(i)})$ is identical to the ground truth posterior distribution $p_\theta(\boldsymbol{Z}|\boldsymbol{X}^{(i)})$, then the $\log p_\theta(\boldsymbol{X}^{(i)}) = \mathcal{L}(\theta, \phi; \boldsymbol{X}^{(i)})$. Therefore, $\mathcal{L}(\theta, \phi; \boldsymbol{X}^{(i)})$ is called the evidence lower bound (ELOB), which is defined by Eq. 2.

$$\begin{aligned}
\mathcal{L}(\theta, \phi; X^{(i)}) &= \log p_\theta(X^{(i)}) - D_{KL}\left(q_\phi(Z|X^{(i)}) \| p_\theta(Z|X^{(i)})\right) \\
&= \mathbb{E}_{q_\phi(Z|X^{(i)})}[\log p_\theta(X^{(i)}|Z)] - D_{KL}\left(q_\phi(Z|X^{(i)}) \| p_\theta(Z|X^{(i)})\right)
\end{aligned} \quad (2)$$

Thus, minimizing the KL divergence is equivalent to maximizing the ELOB. To train the model explicitly and implement the loss function in a closed form, we parameterize the $q_\phi$ as a multivariate normal distribution (multivariate Gaussian distribution) with an approximately diagonal variance-covariance matrix. Then the analytical solution for the KL divergence is shown in Eq. 3.

$$D_{KL}\left(q_\phi(Z|X^{(i)}) \| p_\theta(Z|X^{(i)})\right) = \frac{1}{2}\sum_{d=1}^{D}\left(\left(\mu_d^{(i)}\right)^2 + \left(\sigma_d^{(i)}\right)^2 - \log\left(\left(\sigma_d^{(i)}\right)^2\right) - 1\right) \quad (3)$$

where $D$ is the number of the latent variables extracted by the VAE, and $\mu_d^{(i)}$ and $\left(\sigma_d^{(i)}\right)^2$ are the approximate mean and variance of the posterior distribution of $d$-th latent variable for $i$-th subject.

We extend the VAE from single-view input into multi-view input fashion for multi-view metabolomics data imputation. Notably, as the fact that the product of Gaussian distributions is also a Gaussian distribution [29,30], we apply the Product of the Expert (PoE) to generate the common latent space for the variation inference with an analytical solution. Suppose that under the multi-view setting, we have the data in $V$ views, i.e. $X_1, X_2, \cdots, X_V$. For the data in $v$-th view ($v \in \{1, \cdots, V\}$), a nonlinear function implemented by a neural network is employed as the encoder, denoted as $q_{\phi_v}(Z_v|X_v^{(i)})$, where $\phi_v$ represents the learnable parameters of the nonlinear function for $v$-th view. For each encoder, we estimate the mean vector and the variance-covariance matrix of multivariate Gaussian distribution for the approximate posterior distribution, denoted as $\boldsymbol{\mu}_v^{(i)}$ and $\boldsymbol{\Sigma}_v^{(i)}$ for $i$-th subject, and we assume $\boldsymbol{\mu}_v^{(i)} \in \mathbb{R}^D$ is a vector and $\boldsymbol{\Sigma}_v^{(i)} \in \mathbb{R}^{D \times D}$ is a diagonal matrix. In our implementation, we employ multi-layer perceptron (MLP) as the encoder. To guarantee the positivity of the covariance, the output of the MLP is denoted as the $\log \boldsymbol{\Sigma}_v^{(i)}$ first and then is converted to $\boldsymbol{\Sigma}_v^{(i)}$ using the exponential function. Formally, the encoder is defined in Eq. 4.

$$\begin{aligned}
q_{\phi_v}(Z_v|X_v^{(i)}) &= \mathcal{N}\left(\boldsymbol{\mu}_v^{(i)}, \boldsymbol{\Sigma}_v^{(i)}\right) \\
&= \frac{1}{(2\pi)^{D/2}\sqrt{|\boldsymbol{\Sigma}_v^{(i)}|}} \exp\left(-\frac{1}{2}\left(Z_v - \boldsymbol{\mu}_v^{(i)}\right)^T\left(\boldsymbol{\Sigma}_v^{(i)}\right)^{-1}\left(Z_v - \boldsymbol{\mu}_v^{(i)}\right)\right) \\
\boldsymbol{\mu}_v^{(i)} &= MLP_v^{\mu}\left(X_v^{(i)}\right) \\
\boldsymbol{\Sigma}_v^{(i)} &= \exp\left(MLP_v^{\Sigma}\left(X_v^{(i)}\right)\right)
\end{aligned} \quad (4)$$

where $Z_v \in \mathbb{R}^D$ is the latent variable extracted by $v$-th view with the dimension of $D$. $MLP_v^{\mu}$ and $MLP_v^{\Sigma}$ are the neural networks for calculating mean and covariance for the Gaussian distribution, respectively. Let $\mathbf{T}_v^{(i)} = \left(\boldsymbol{\Sigma}_v^{(i)}\right)^{-1}$, then the multivariate Gaussian distribution for $v$-th view is rewritten as Eq. 5.

$$q_{\phi_v}(Z_v|X_v^{(i)}) = \frac{1}{(2\pi)^{D/2}\sqrt{|\boldsymbol{\Sigma}_v^{(i)}|}} \exp\left(-\frac{1}{2}Z_v^T \mathbf{T}_v^{(i)} Z_v + \left(\boldsymbol{\mu}_v^{(i)}\right)^T \mathbf{T}_v^{(i)} Z_v + \Delta_v^{(i)}\right) \quad (5)$$

where $\Delta_v^{(i)} = -\frac{1}{2}\left(\boldsymbol{\mu}_v^{(i)}\right)^T \mathbf{T}_v^{(i)} \boldsymbol{\mu}_v^{(i)} - \frac{D}{2}\log 2\pi + \frac{1}{2}\log\left|\mathbf{T}_v^{(i)}\right|$. A PoE modeled the target posterior distribution of the common latent variable from multi-view as the product of the individual posterior distribution of the latent variable from single-view. According to Eq. 5, $\Delta_v^{(i)}$ was not related to the latent variable $\mathbf{Z}_v$. Therefore, for the following analysis, $\Delta_v^{(i)}$ was considered as a constant. As a result, the PoE generated the common latent variable $\mathbf{Z}$, which was defined in Eq. 6.

$$q_\phi\left(\mathbf{Z}\big|\mathbf{X}_1^{(i)}\cdots\mathbf{X}_V^{(i)}\right) = \frac{1}{V}\prod_{v=1}^{V} q_{\phi_v}\left(\mathbf{Z}_v\big|\mathbf{X}_v^{(i)}\right) \tag{6}$$

Eq.6 indicated that the multivariate Gaussian distribution of the common latent variable was defined by the product of the multivariate Gaussian distribution of the latent variable extracted by $V$ views. According to the approximated posterior distribution of the common latent variable, $\mathbf{Z}$, was derived in Eq. 7.

$$\begin{aligned}
q_\phi\left(\mathbf{Z}\big|\mathbf{X}_1^{(i)}\cdots\mathbf{X}_V^{(i)}\right) &= \mathcal{N}\left(\boldsymbol{\mu}_z^{(i)}, \boldsymbol{\Sigma}_z^{(i)}\right), \\
\boldsymbol{\mu}_z^{(i)} &= \left(\sum_{v=1}^{V}\left(\boldsymbol{\mu}_v^{(i)}\right)^T \mathbf{T}_v^{(i)}\right)\left(\sum_{v=1}^{V}\mathbf{T}_v^{(i)}\right)^{-1} \\
\boldsymbol{\Sigma}_z^{(i)} &= \left(\sum_{v=1}^{V}\mathbf{T}_v^{(i)}\right)^{-1}
\end{aligned} \tag{7}$$

where $\boldsymbol{\mu}_z^{(i)}$ and $\boldsymbol{\Sigma}_z^{(i)}$ were the mean vector and variance-covariance matrix of the approximated posterior distribution of common latent variable for the $i$-th subject. To make the neural network differentiable, we adopted the reparameterization trick [28] to reparametrize the mean vector and the diagonal variance-covariance matrix of the multi-variate Gaussian distribution, as shown in Eq. 8.

$$\mathbf{Z}^{(i)} = \boldsymbol{\mu}_z^{(i)} + \left(\boldsymbol{\Sigma}_z^{(i)}\right)^{1/2} \odot \boldsymbol{\epsilon}_z \tag{8}$$

where $\boldsymbol{\epsilon}_z \sim \mathcal{N}(0, \mathbf{I})$ and $\odot$ indicates the element-wise product. Similar to the architecture of the encoder, we employed MLPs as the decoder to restore the integrated features, denoted as $f_v^{dec}$ for the $v$-th view. Formally, the reconstructed features for the $v$-th view was denoted as $\widehat{\mathbf{X}}_v^{(i)}$, and the reconstruction was defined in Eq. 9.

$$\widehat{\mathbf{X}}_v^{(i)} = MLP_v^{dec}\left(\mathbf{Z}^{(i)}\right) \tag{9}$$

### 2.4. Loss function and model training

20% of the subjects were randomly chosen as the test set, and the rest of the data were used as the training set. Since the product of the Gaussian distributions was another Gaussian distribution, we employed the ELOB designed for variational autoencoder with the explicit form as the objective function to optimize the neural network, as shown in Eq. 10.

$$\begin{aligned}
\mathcal{L}(\theta, \phi; \mathbf{X}_1, \cdots, \mathbf{X}_V) &= \sum_{i=1}^{N}\sum_{v=1}^{V} \mathbb{E}_{\mathbf{Z} \sim q_\phi\left(\mathbf{Z}^{(i)}\big|\mathbf{X}_v^{(i)}\right)} \log p_\theta\left(\mathbf{X}_v^{(i)}\big|\mathbf{Z}\right) \\
&- \sum_{i}^{N} D_{KL}\left(q_\phi\left(\mathbf{Z}^{(i)}\big|\mathbf{X}_1^{(i)}, \cdots, \mathbf{X}_V^{(i)}\right) \Vert p_\theta\left(\mathbf{Z}^{(i)}\big|\mathbf{X}_1^{(i)}, \cdots, \mathbf{X}_V^{(i)}\right)\right)
\end{aligned} \tag{10}$$

As shown in Eq. 10, the ELOB contained two terms, where the first term on the right hand side of Eq. 10 penalized the discrepancy between the reconstructed features and the input feature and the second term on the right hand side measured the KL-divergence between the prior and posterior distributions. The analytical form of the KL-divergence was derived according to VAE [31] and the overall loss function was shown in Eq. 11.

$$\mathcal{L}(\theta, \phi; \mathbf{X}_1, \cdots, \mathbf{X}_V) = \sum_{i=1}^{N} \sum_{v=1}^{V} \left( X_v^{(i)} \log\left(\hat{X}_v^{(i)}\right) + \left(1 - X_v^{(i)}\right) \log\left(1 - \hat{X}_v^{(i)}\right) \right) \\ - \left( \frac{1}{2} \sum_{i=1}^{N} \sum_{d=1}^{D} \left( \left(\mu_d^{(i)}\right)^2 + \left(\sigma_d^{(i)}\right)^2 - \log\left(\sigma_d^{(i)}\right)^2 - 1 \right) \right) \quad (11)$$

where $\mu_d^{(i)}$ and $\left(\sigma_d^{(i)}\right)^2$ were the approximate mean and variance of the posterior distribution of the $d$-th latent variable for the $i$-th subject.

Since the burden score was generated according to the $M$ template, we designed an algorithm to train the MVAE based on prior knowledge, as shown in Algorithm 1.

**Algorithm 1**. MVAE training algorithm for metabolomics data imputation

> **Input**:
> - $X_{tr} = \{\mathbf{X}_{PGS}^{tr}, \mathbf{X}_{LD}^{tr}\}$, representing PGS and LD pruned SNPs for the subjects in the training set
> - $Y_j = \{\mathbf{y}^{j^{tr}}, \mathbf{y}^{j^{te}}\}$, representing **one** metabolite indexed by $j$ for the subjects in both the training set and testing set
> - $M$: template size
> - $X_{BS}^{Template} = \{(\mathbf{X}_{BS}^{1^{tr}}, \mathbf{X}_{BS}^{1^{te}}), (\mathbf{X}_{BS}^{2^{tr}}, \mathbf{X}_{BS}^{2^{te}}), \cdots, (\mathbf{X}_{BS}^{M^{tr}}, \mathbf{X}_{BS}^{M^{te}})\}$: burden scores for $M$ metabolites in template set
> - $Y^{Template} = \{(\mathbf{y}^{1^{tr}}, \mathbf{y}^{1^{te}}), (\mathbf{y}^{2^{tr}}, \mathbf{y}^{2^{te}}), \cdots, (\mathbf{y}^{M^{tr}}, \mathbf{y}^{M^{te}})\}$: metabolites corresponding to $M$ metabolites in $X_{BS}^{Template}$
>
> **Output**: trained MVAE for $j$-th metabolite
>
> **Training**:
> - Calculate the Pearson correlation between $\mathbf{y}^{j^{tr}}$ and each $\mathbf{y}^{i^{tr}}$ in template set $Y^{Template}$.
> - Select the template metabolite with the highest Pearson correlation with index $i, s.t. i \in \{1, \cdots, M\}$.
> - Copy burden score $\mathbf{X}_{BS}^{i^{tr}}$ from template set $X_{BS}^{Template}$ to generate multi-view dataset $X_{tr} = \{\mathbf{X}_{BS}^{i^{tr}}, \mathbf{X}_{PGS}^{tr}, \mathbf{X}_{LD}^{tr}\}$
> - Train MVAE with the input of $X_{tr}$ and $\mathbf{y}^{j^{tr}}$ using loss function defined in Eq. 11.

In algorithm 1, the PGS and the LD-pruned SNPs were generated according to the WGS data; while the burden scores were calculated by the residual of the linear regression model trained using the PCs and the template metabolites. The designed algorithm assumed that the selected template metabolites exist in both the training subjects and the testing subjects. Thus, the burden scores for the template metabolites were presented in both the training subjects and the testing subjects. The designed algorithm was for feature-level metabolomics data imputation. The MVAE was built for training and prediction for one specific metabolite. To impute the missing metabolites in the dataset, multiple MVAEs were required to be trained.

In algorithm 1, the template set contained $M$ metabolites with corresponding PGS and LD-pruned SNPs. For each predicted metabolite, the Pearson correlation between $y_{tr}$ and each $y^{i^{tr}}$ in the metabolite template set was compared. The template metabolite with the highest Pearson correlation was selected and the burden score of the template metabolite was copied to form the training set $X_{tr} = \{\mathbf{X}_{BS}^{i^{tr}}, \mathbf{X}_{PGS}^{tr}, \mathbf{X}_{LD}^{tr}\}$. Using the training data $X_{tr} = \{\mathbf{X}_{BS}^{i^{tr}}, \mathbf{X}_{PGS}^{tr}, \mathbf{X}_{LD}^{tr}\}$ and the corresponding metabolite $y_{tr}$, the MVAE was built. During the testing, the template metabolite was used to generate the burden score $\mathbf{X}_{BS}^{ite}$ from the template set $X_{BS}^{Template}$, and then to generate the multi-view dataset $X_{te} = \{\mathbf{X}_{BS}^{ite}, \mathbf{X}_{PGS}^{te}, \mathbf{X}_{LD}^{te}\}$. Using the trained MVAE, the metabolite was imputed. The testing algorithm is shown in Algorithm 2.

**Algorithm 2**. MVAE testing algorithm for metabolomics data imputation.

---
**Input**:
- $X_{te} = \{\mathbf{X}_{PGS}^{te}, \mathbf{X}_{LD}^{te}\}$, representing PGS and LD pruned SNPs for the subjects in the testing set
- $M$: template size
- $X_{BS}^{Template} = \{(\mathbf{X}_{BS}^{1tr}, \mathbf{X}_{BS}^{1te}), (\mathbf{X}_{BS}^{2tr}, \mathbf{X}_{BS}^{2te}), \cdots, (\mathbf{X}_{BS}^{Mtr}, \mathbf{X}_{BS}^{Mte})\}$: burden scores for $M$ metabolites in template set
- $Y^{Template} = \{(\mathbf{y}^{1tr}, \mathbf{y}^{1te}), (\mathbf{y}^{2tr}, \mathbf{y}^{2te}), \cdots, (\mathbf{y}^{Mtr}, \mathbf{y}^{Mte})\}$: metabolites corresponding to $M$ metabolites in $X_{BS}^{Template}$
- Trained MVAE for $j$-th metabolite

**Output**: $\mathbf{y}_{pred}^{jte}$: model imputed metabolite for $j$-th metabolite in testing set

**Testing**:
- Calculate the Pearson correlation between $\mathbf{y}^{jtr}$ and each $y^{itr}$ in template set $Y^{Template}$.
- Select the template metabolite with the highest Pearson correlation with index $i, s.t. i \in [1, M]$
- Copy burden score $\mathbf{X}_{BS}^{ite}$ from template set $X_{BS}^{Template}$ to generate multi-view dataset $X_{te} = \{\mathbf{X}_{BS}^{ite}, \mathbf{X}_{PGS}, \mathbf{X}_{LD}\}$
- Use the trained MVAE to predict $\mathbf{y}_{pred}^{jte}$ according to $X_{te} = \{\mathbf{X}_{BS}^{ite}, \mathbf{X}_{PGS}, \mathbf{X}_{LD}\}$

---

### 2.5. Model evaluation

For model evaluation, mean absolute percentage error (MAPE) and $R^2$-score were employed. A lower MAPE and a higher $R^2$-score indicate better performance. 0 of MAPE indicates the perfect match. $R^2$-score ranges from $-\infty$ to 1, where 1 indicates the perfect match.

### 3. Results and discussion

### 3.1. Model performance for metabolomics data imputation

Our designed MVAE model was implemented using TensorFlow 2.5. We performed grid search to find the optimal neural network architecture. In our implementation, the used MLPs, including $MLP_v^\mu$, $MLP_v^\Sigma$ and $MLP_v^{dec}$ contained 2 fully connected layers with 128 neurons. The distribution for each view is a 64-dimensional Gaussian distribution and the product of these Gaussian distribution has the dimension of 64, i.e. $\mathbf{Z}^{(i)} \in \mathbb{R}^{64}$. To generate an overall performance comparison, we set the cut-off thresholds of 0.01, 0.05, 0.1, 0.15 and 0.2 for the overall $R^2$-scores to measure performance of metabolite imputation; similarly, we determined the thresholds of 0.1, 0.15, 0.2 and 0.3 for the overall MAPEs to measure the performance of metabolite imputation. All the subsequent results were based on independent testing data.

In our dataset, 497 metabolites were enrolled. We tested the model performance with different number of metabolites as the templates. Then the rest $497 - M$ metabolites were used to train MVAE and evaluate the model performance. We depicted the performance of the proposed MVAE in Figure 2 and the detailed performance is shown in Table S1 to S4 in the supplementary materials.

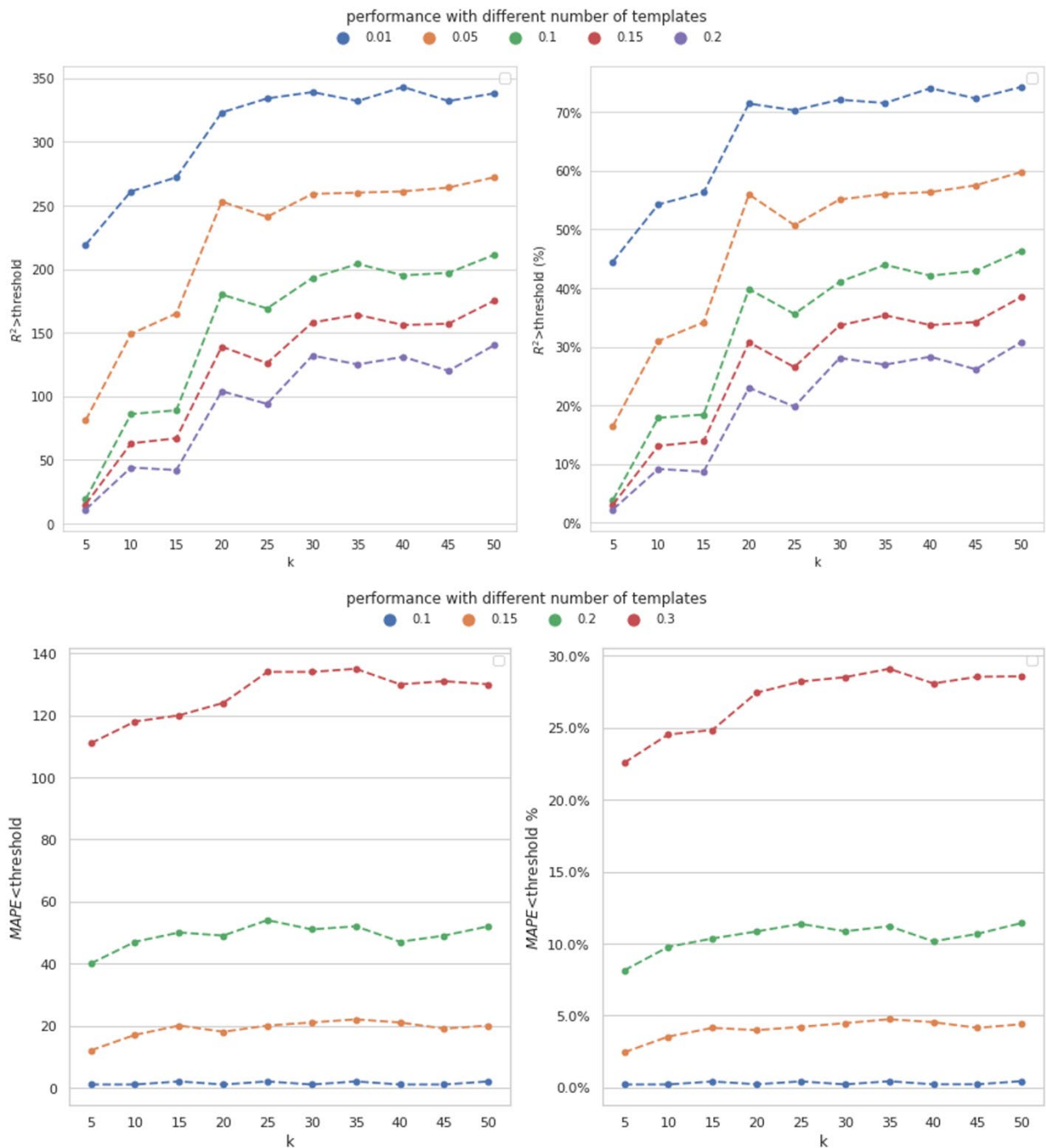

**Figure 2**. Performance of metabolomics imputation using different number of templates by MVAE. Distinct colors indicate the performance achieved across varying thresholds. Top: the achieved $R^2$-score using different number of template metabolites under different thresholds; Bottom: the achieved MAPE using different number of template metabolites under different thresholds. The count of metabolites achieved the corresponding performance, and the percent of metabolites are depicted.

In Figure 2, the analysis of the model performance with respect to the number of template metabolites reveals interesting trends. As the number of template metabolites (M) increases up to 30, the model's performance consistently improves, indicating that the inclusion of more template metabolites enhances the accuracy of the imputation. This suggests that a larger set of template metabolites provides more comprehensive information, enabling the model to make more accurate predictions.

However, beyond a certain point ($M > 35$), the trend changed, and the model's performance did not exhibit consistent improvement. This observation suggests that there might be a saturation point, after which adding more template metabolites does not significantly contribute to enhancing the model's accuracy. The results indicate that our model is robust, as it demonstrates the capacity to impute all metabolites enrolled in our dataset with only 7.04% (35/497) of known metabolites. This finding highlights the effectiveness of our approach in handling missing metabolomics data. Despite having access to only a small portion of known metabolites, our model is capable of accurately predicting and imputing the entire set of metabolites, making it a promising and practical solution for missing value imputation in mass spectrometry-based metabolomics data.

### 3.2. Model performance comparison

To illustrate the effectiveness of the designed MVAE, three multi-view integration algorithms were enrolled, including:

- Multiview canonical correlation analysis (MCCA) [32]. MCCA extends the canonical correlation analysis (CCA) into multi-view settings. CCA is a typical subspace learning algorithm, aiming at finding the pairs of projections from different views with the maximum correlations. For more than 2 views, MCCA optimizes the sum of pairwise correlations.
- Kernel CCA (KCCA) [33]. KCCA is based on MCCA, however, it adds a centered Gram matrix to perform the nonlinear transformation on the input data.
- Kernel generalized CCA (KGCCA) [34]. KGCCA extends KCCA with a priori-defined graph connections between different views.

In addition, one of the novel aspects of this study is the cross-omics imputation, which incorporates both WGS data and template metabolites. To further highlight the effectiveness of our approach, we conducted a comparison by solely using metabolomics data for within-omics imputation. Specifically, the within-omics imputation involves the model utilizing the template metabolites to perform the imputation. We employed various compared models, including KNN, Ridge regression, support vector machine (SVM), RF regression, and gradient boosting regression (GBT). Each of these models used one template metabolite with the highest Pearson correlation with the imputed metabolite as input to impute the metabolite for the test subjects. Multiple compared models were constructed for different metabolites, and we evaluated the model performance using the MAPE and $R^2$-score with the same cut-off thresholds as mentioned in Section 3.1. The number of template metabolites was fixed at $M = 35$, and the comparison results are presented in Figure 3. The detailed performance is shown in Table S5 to S8 in the supplementary materials.

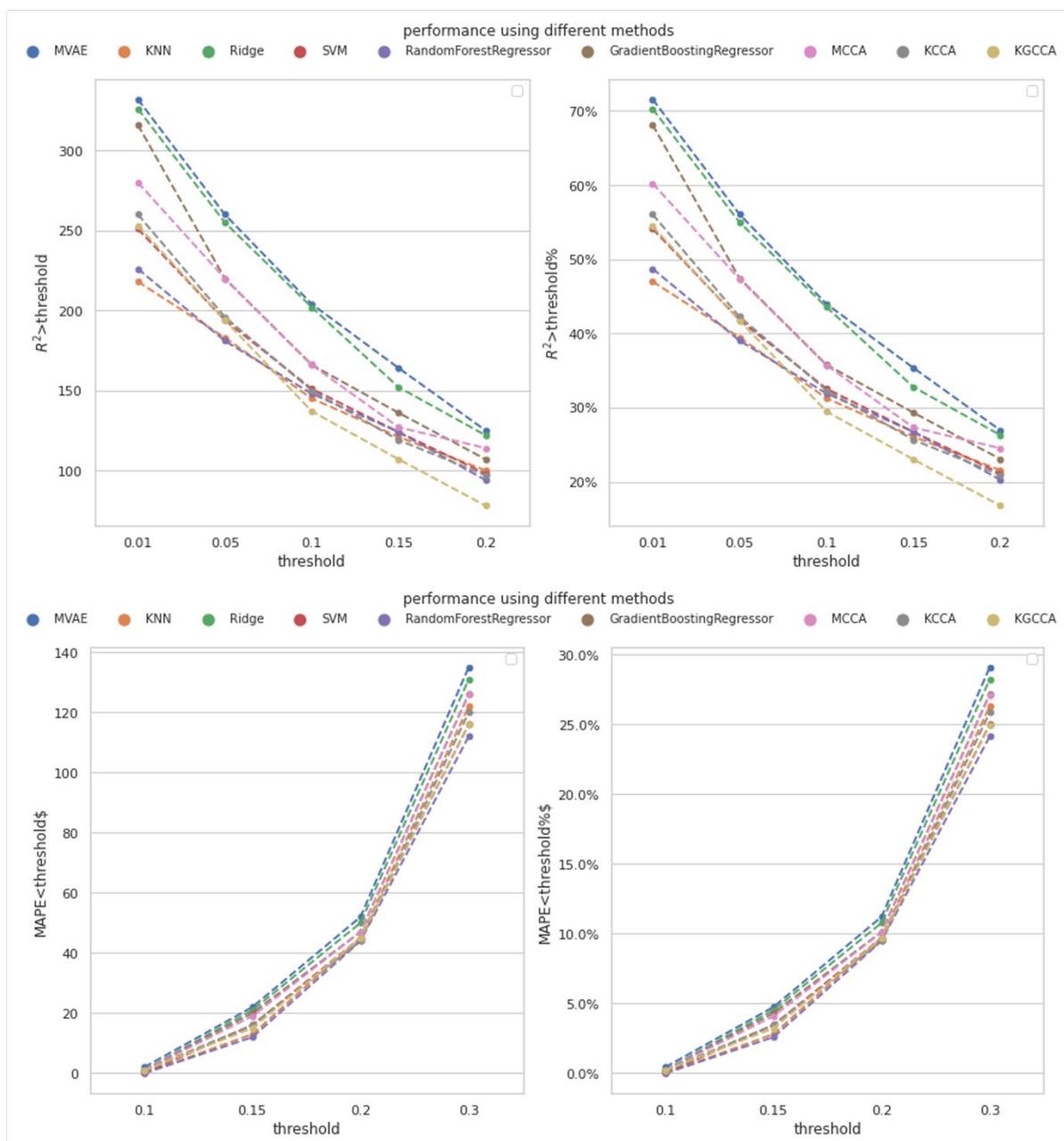

**Figure 3**. Performance of metabolomics imputation using different methods by the proposed MVAE. Top: the achieved $R^2$-score using different numbers of template metabolites under different thresholds; Bottom; the achieved MAPE using different number of template metabolites under different thresholds. The count of metabolites achieved the corresponding performance, and the percent of metabolites are depicted. The pseudo colors indicate different models.

According to Figure 3, the designed MVAE model achieved the highest performance because the number of the metabolites with satisfactory $R^2$-score was consistently higher than the compared methods. Regarding MAPE, the proposed MVAE achieved imputed metabolites with approximately 30% having a MAPE smaller than 0.3. This outcome is a strong indicator of the robustness and superior performance of the MVAE approach in effectively modeling high-dimensional multi-view genomics data.

As an increasing number of studies integrate WGS data with metabolomics data to gain a comprehensive understanding of the fundamental molecular underpinnings of biological processes and diseases, few methods are available to perform cross-omics-based imputation. MVAE leverages the power of variational autoencoders to learn the underlying latent representations from multi-view genomics data, enabling it to capture complex relationships and dependencies among different omics modalities. This ability to jointly model diverse genomic features contributes to the improved accuracy and reliability of the imputation process. Furthermore, MVAE demonstrates its adaptability to handle high-dimensional data, which is a common challenge in genomics research. By efficiently extracting relevant information from multiple views, MVAE effectively overcomes the curse of dimensionality and provides more accurate imputations even with limited information.

We also conducted an in-depth evaluation against within-omics imputation methods, including SVM, Ridge regression, random forest, gradient descent boosting, and KNN. The results demonstrated that MVAE consistently outperformed these within-omics imputation methods in terms of both $R^2$-scores and MAPEs. This superior performance of the cross-omics approach can be attributed to MVAE's ability to effectively leverage information from multiple views, which enhances its imputation accuracy and robustness. The within-omics imputation methods, on the other hand, rely solely on one template metabolite with the highest Pearson correlation, making them more limited in capturing the complex relationships and dependencies present in multi-view genomics data.

The higher $R^2$-scores obtained by MVAE further emphasize its capacity to produce more accurate imputations, surpassing the performance of traditional within-omics imputation techniques. This improvement in imputation accuracy is of paramount importance in various applications, such as gene expression prediction, functional annotation, and pathway analysis, where precise imputation plays a crucial role in obtaining reliable downstream analysis results. Moreover, MVAE's ability to achieve better MAPEs highlights its efficiency in imputing metabolites with a high degree of precision, ensuring minimal errors in the imputed values. Its ability to outperform traditional within-omics imputation methods and achieve superior $R^2$-scores and MAPEs reaffirms its potential to revolutionize the field of multi-view genomics data integration and analysis. This is particularly significant in genomics research, as it reduces the impact of missing data on downstream analyses, leading to more robust and reliable interpretations.

In summary, our results highlight the effectiveness of the proposed MVAE model as a powerful tool for modeling high-dimensional multi-view genomics data. Its ability to achieve superior $R^2$-scores compared to existing methods emphasizes its potential in addressing the challenges of data integration and imputation in the field of genomics research.

### 4. Conclusion

In this paper, we addressed the common challenge of missing data in mass spectrometry-based metabolomics data imputation by proposing a novel and effective multi-view information fusion method. We presented an MVAE framework to integrate common/rare variants and template metabolites for joint feature extraction and cross-omics data imputation. By learning latent representations from both omics data, our approach demonstrated superior imputation performance compared to conventional techniques. Our method achieved remarkable accuracy in imputing missing metabolomics values, with a significant $R^2$-score (> 0.01) for 72.13% of metabolites, using only 35 template metabolites. These results underscored the potential of our approach to improve data completeness and enhance multi-omics integration studies. Overall, our proposed method showcased the benefits of combining WGS data with metabolomics in data imputation, paving the way for more comprehensive and accurate investigations in the fields of metabolomics and precision medicine.

## Acknowledgments

This research was supported in part by grants from the National Institutes of Health, USA (P20GM109036, R01AR069055, U19AG055373, R01AG061917, and R15HL172198) and NASA Johnson Space Center, USA contracts NNJ12HC91P and NNJ15HP23P. It was also supported in part by seed grants from the Michigan Technological University Institute of Computing and Cybersystems, a graduate fellowship from Michigan Technological University Health Research Institute, and a graduate fellowship from Portage Health Foundation.

**Supplementary materials**.

Table S1. Performance of metabolomics imputation using different number of templates by the proposed MVAE. The count of imputed metabolites achieved $R^2$-scores greater than thresholds are presented.

| k | 5 | 10 | 15 | 20 | 25 | 30 | 35 | 40 | 45 | 50 |
|---|---|---|---|---|---|---|---|---|---|---|
| 0.01 | 219 | 261 | 272 | 323 | 334 | 339 | 332 | 343 | 332 | 338 |
| 0.05 | 81 | 149 | 165 | 253 | 241 | 259 | 260 | 261 | 264 | 272 |
| 0.1 | 19 | 86 | 89 | 180 | 169 | 193 | 204 | 195 | 197 | 211 |
| 0.15 | 15 | 63 | 67 | 139 | 126 | 158 | 164 | 156 | 157 | 175 |
| 0.2 | 11 | 44 | 42 | 104 | 94 | 132 | 125 | 131 | 120 | 140 |

Table S2. Performance of metabolomics imputation using different number of templates by the proposed MVAE. The percent of imputed metabolites achieved $R^2$-scores greater than thresholds are presented.

| k | 5 | 10 | 15 | 20 | 25 | 30 | 35 | 40 | 45 | 50 |
|---|---|---|---|---|---|---|---|---|---|---|
| 0.01 | 44.51 | 54.26 | 56.31 | 71.46 | 70.32 | 72.13 | 71.55 | 74.08 | 72.33 | 74.29 |
| 0.05 | 16.46 | 30.98 | 34.16 | 55.97 | 50.74 | 55.11 | 56.03 | 56.37 | 57.52 | 59.78 |
| 0.1 | 3.86 | 17.88 | 18.43 | 39.82 | 35.58 | 41.06 | 43.97 | 42.12 | 42.92 | 46.37 |
| 0.15 | 3.05 | 13.10 | 13.87 | 30.75 | 26.53 | 33.62 | 35.34 | 33.69 | 34.20 | 38.46 |
| 0.2 | 2.24 | 9.15 | 8.70 | 23.01 | 19.79 | 28.09 | 26.94 | 28.29 | 26.14 | 30.77 |

Table S3. Performance of metabolomics imputation using different number of templates by the proposed MVAE. The count of imputed metabolites achieved MAPEs smaller than thresholds are presented.

| k | 5 | 10 | 15 | 20 | 25 | 30 | 35 | 40 | 45 | 50 |
|---|---|---|---|---|---|---|---|---|---|---|
| 0.1 | 1 | 1 | 2 | 1 | 2 | 1 | 2 | 1 | 1 | 2 |
| 0.15 | 12 | 17 | 20 | 18 | 20 | 21 | 22 | 21 | 19 | 20 |
| 0.2 | 40 | 47 | 50 | 49 | 54 | 51 | 52 | 47 | 49 | 52 |
| 0.3 | 111 | 118 | 120 | 124 | 134 | 134 | 135 | 130 | 131 | 130 |

Table S4. Performance of metabolomics imputation using different number of templates by the proposed MVAE. The percent of imputed metabolites achieved MAPEs smaller than thresholds are presented.

| k | 5 | 10 | 15 | 20 | 25 | 30 | 35 | 40 | 45 | 50 |
|---|---|---|---|---|---|---|---|---|---|---|
| 0.1 | 0.20 | 0.21 | 0.41 | 0.22 | 0.42 | 0.21 | 0.43 | 0.22 | 0.22 | 0.44 |
| 0.15 | 2.44 | 3.53 | 4.14 | 3.98 | 4.21 | 4.47 | 4.74 | 4.54 | 4.14 | 4.40 |
| 0.2 | 8.13 | 9.77 | 10.35 | 10.84 | 11.37 | 10.85 | 11.21 | 10.15 | 10.68 | 11.43 |
| 0.3 | 22.56 | 24.53 | 24.84 | 27.43 | 28.21 | 28.51 | 29.09 | 28.08 | 28.54 | 28.57 |

**Table S5**. Performance of metabolomics imputation using different methods with 35 template metabolites. The count of imputed metabolites achieved $R^2$-scores greater than thresholds are presented.

| method | MVAE | KNN | Ridge | SVM | RF | GBR | MCCA | KCCA | KGCCA |
|---|---|---|---|---|---|---|---|---|---|
| 0.01 | 332 | 218 | 326 | 251 | 226 | 316 | 280 | 260 | 253 |
| 0.05 | 260 | 183 | 255 | 194 | 181 | 220 | 220 | 196 | 194 |
| 0.1 | 204 | 145 | 202 | 151 | 148 | 166 | 166 | 150 | 137 |
| 0.15 | 164 | 121 | 152 | 124 | 124 | 136 | 127 | 119 | 107 |
| 0.2 | 125 | 100 | 122 | 98 | 94 | 107 | 114 | 97 | 78 |

**Table S6**. Performance of metabolomics imputation using different methods with 35 template metabolites. The percent of imputed metabolites achieved $R^2$-scores greater than thresholds are presented.

| method | MVAE | KNN | Ridge | SVM | RF | GBR | MCCA | KCCA | KGCCA |
|---|---|---|---|---|---|---|---|---|---|
| 0.01 | 71.55 | 46.98 | 70.26 | 54.09 | 48.71 | 68.10 | 60.22 | 56.03 | 54.41 |
| 0.05 | 56.03 | 39.44 | 54.96 | 41.81 | 39.01 | 47.41 | 47.31 | 42.24 | 41.72 |
| 0.1 | 43.97 | 31.25 | 43.53 | 32.54 | 31.90 | 35.78 | 35.70 | 32.33 | 29.46 |
| 0.15 | 35.34 | 26.08 | 32.76 | 26.72 | 26.72 | 29.31 | 27.31 | 25.65 | 23.01 |
| 0.2 | 26.94 | 21.55 | 26.29 | 21.12 | 20.26 | 23.06 | 24.52 | 20.91 | 16.77 |

**Table S7**. Performance of metabolomics imputation using different methods with 35 template metabolites. The count of imputed metabolites achieved MAPE smaller than thresholds are presented.

| method | MVAE | KNN | Ridge | SVM | RF | GBR | MCCA | KCCA | KGCCA |
|---|---|---|---|---|---|---|---|---|---|
| 0.1 | 2 | 0 | 1 | 0 | 0 | 1 | 1 | 1 | 1 |
| 0.15 | 22 | 13 | 21 | 16 | 12 | 20 | 19 | 16 | 15 |
| 0.2 | 52 | 45 | 50 | 45 | 44 | 47 | 47 | 44 | 45 |
| 0.3 | 135 | 122 | 131 | 116 | 112 | 126 | 126 | 120 | 116 |

**Table S8**. Performance of metabolomics imputation using different methods with 35 template metabolites. The percent of imputed metabolites achieved MAPE smaller than thresholds are presented.

| method | MVAE | KNN | Ridge | SVM | RF | GBR | MCCA | KCCA | KGCCA |
|---|---|---|---|---|---|---|---|---|---|
| 0.1 | 0.43 | 0.00 | 0.22 | 0.00 | 0.00 | 0.22 | 0.22 | 0.22 | 0.22 |
| 0.15 | 4.74 | 2.80 | 4.53 | 3.45 | 2.59 | 4.31 | 4.09 | 3.45 | 3.23 |
| 0.2 | 11.21 | 9.70 | 10.78 | 9.70 | 9.48 | 10.13 | 10.11 | 9.48 | 9.68 |
| 0.3 | 29.09 | 26.29 | 28.23 | 25.00 | 24.14 | 27.16 | 27.10 | 25.86 | 24.95 |